\documentclass[aps,twocolumn,superscriptaddress]{revtex4-1}
\usepackage{graphicx}
\usepackage{amsmath}
\usepackage{amssymb}
\usepackage{booktabs}
\usepackage[colorlinks=True,linkcolor=blue,anchorcolor=blue,citecolor=blue,CJKbookmarks=true]{hyperref}

\usepackage{ulem}

\begin{document}
\title{Electron-hole pair production in graphene for two arbitrarily polarized electric fields with a time delay}

\author{R. Z. Jiang}
\affiliation{State Key Laboratory for Tunnel Engineering, China University of Mining and Technology, Beijing 100083, China}

\author{Z. L. Li}
\email{Corresponding author. zlli@cumtb.edu.cn}
\affiliation{State Key Laboratory for Tunnel Engineering, China University of Mining and Technology, Beijing 100083, China}
\affiliation{School of Science, China University of Mining and Technology, Beijing 100083, China}

\author{Y. J. Li}
\email{Corresponding author. lyj@aphy.iphy.ac.cn}
\affiliation{State Key Laboratory for Tunnel Engineering, China University of Mining and Technology, Beijing 100083, China}
\affiliation{School of Science, China University of Mining and Technology, Beijing 100083, China}

\date{\today}

\begin{abstract}
The momentum distributions of electron-hole (EH) pair production in graphene for two arbitrarily polarized electric fields with a time delay are investigated employing a massless quantum kinetic equation and compared with the results obtained in electron-positron (EP) pair production from vacuum.
For a single elliptically polarized electric field, the momentum distributions of created EH and EP pairs are similar in multiphoton absorption region.
However, for two co-directional linearly polarized electric fields with a time delay and no field frequency, the momentum distribution of created EH pairs exhibits ring patterns, which is not present in EP pair production.
For two circularly polarized fields with identical or opposite handedness, the momentum distributions of created EH pairs also show Ramsey interference and spiral structures, respectively.
Different from EP pair production, the spiral structures are insensitive to the number of oscillation cycles in electric field pulses.
For two elliptically polarized fields with same-sign or opposite-sign ellipticity, the momentum distributions of EH pairs are much more insensitive to ellipticity than those in EP pair production.
These results provide further theoretical reference for simulating the EP pair production from vacuum in solid-state systems.
\end{abstract}

\maketitle

\section{INTRODUCTION}
\label{sec:one}
The vacuum can produce electron-positron (EP) pairs by tunneling under the condition of strong electromagnetic fields, which was first predicted by Sauter in $1931$ \cite{Sauter1931}.
And then, Schwinger \cite{Schwinger1951} calculated the production rate of EP pairs, $\mathcal{P}\sim\exp\left(-\pi E_{\mathrm{cr}}/E_0\right)$, in a constant electric field using the proper-time method.
Here $E_{\mathrm{cr}}=m^2c^3/e\hbar \approx 1.3\times 10^{18}\mathrm{V}/\mathrm{m}$ is the critical electric field strength, $E_0$ is the external electric field strength, $m$ and $-e$ are the mass and charge of electrons, respectively.
The laser intensity corresponding to the critical field strength $E_{\mathrm{cr}}$ is about $10^{29}\mathrm{W}/\mathrm{cm}^2$, which is several orders of magnitude higher than the currently achievable one $10^{23}\mathrm{W}/\mathrm{cm}^2$ \cite{Yoon2021}.
Because of the exponential suppression of the production rate, it is still not possible to observe this phenomenon in the laboratory.
Therefore, many ways to increase the particle pair yield have been proposed, such as by using frequency-chirped electric fields \cite{Abdukerim2017,Gong2020,Mohamedsedik2021} and the dynamically assisted Schwinger mechanism \cite{Schutzhold2008,Li2014a,Sitiwaldi2018,Mahlin2023}.
In addition, there is also interest in using solid-state systems to study pair production. One of the systems is graphene.

Graphene is a $2$-dimensional material consisting of a single layer of carbon atoms arranged in a hexagonal honeycomb. Its concept was first introduced by Wallace in 1946 \cite{Wallace1947} when he studied the energy band properties of graphite, and it was first separated from graphite in 2004 by a technique named micromechanical cleavage \cite{Novoselov2004,Novoselov2005}.
Because of its unique electronic properties, graphene is gradually becoming a bridge between condensed matter physics and quantum electrodynamics (QED) \cite{Katsnelson2007}.
Recently, many phenomena similar to the QED process have been found in graphene, such as Klein paradox \cite{Katsnelson2006, Stander2009, Gutierrez2016}, Coulomb supercriticality \cite{Terekhov2008} and Lamb shift \cite{Kibis2011}.
Furthermore, since graphene can produce electron-hole (EH) pairs in an external electric field, it can also be used as a simulation site for EP pair production processes in a vacuum.
In order to make an analogy between the production of EH pairs and EP pairs, the minimal set of experimental conditions is given in Ref. \cite{Fillion-Gourdeau2016}.
There have been many methods to study the excitation of quasi-particles from graphene in the presence of an electric field.
For example, the low-energy approximation model \cite{Smolyansky2019,Aleksandrov2020,Gavrilov2020,Smolyansky2020}, which produces the equations analogous to the quantum Vlasov equation (QVE) \cite{Kluger1998,Schmidt1998,Bloch1999}, is convenient for studying the momentum distribution of electrons created in graphene for a homogeneous electric field.
The quasi-particle quantization method was used to study the momentum spectrum of electrons created by tunneling process and multi-photon process in Ref. \cite{Fillion-Gourdeau2015}.

Nevertheless, the difference between the production of EH pairs in graphene and EP pairs in vacuum for a complex external electric field needs further investigation.
We know that many studies have been done on EP pair production for different electric field configurations (e.g., elliptically polarized electric fields or two electric fields with a time delay), and several fascinating physical phenomena have been discovered.
For instance, the shell structure \cite{Otto2015,Olugh2015}, the time-domain multiple-slit interference \cite{Akkermans2011,Li2014xga}, and the spiral structure \cite{Li2017qwd,Hu2023pmz} in momentum spectra of EP pairs.
Therefore, the signatures of momentum spectra of EH pairs for two arbitrarily polarized electric fields with a time delay attract our research interest.
In this paper, we use the low-energy approximation model to study the following issues: (a) the momentum spectrum of created EH pairs for a single elliptically polarized electric field; (b) the time-domain interference phenomenon in momentum spectrum for two linearly polarized electric fields with a time delay; (c) the effect of the time delay and the number of oscillating periods of the electric field on the momentum spectrum for two circularly polarized electric fields with a time delay; and (d) the effect of ellipticity on the momentum spectrum for two elliptically polarized electric fields with a time delay.

The paper is organized as follows.
In Sec. \ref{sec:two}, a brief description of the low-energy approximation model and the external electric field we used are provided.
In Sec. \ref{sec:three}, the momentum spectrum of EH pairs in the presence of a single elliptically polarized electric field is discussed.
In Sec. \ref{sec:four}, the momentum spectrum produced by two electric fields with a time delay is considered in four cases: two linearly polarized electric fields, two circularly polarized electric fields with identical handedness, two circularly polarized electric fields with opposite handedness, and two elliptically polarized electric fields. These results are discussed in Sec. \ref{A}, Sec. \ref{B}, Sec. \ref{C}, and Sec. \ref{D}, respectively.
Section \ref{sec:five} is a summary and discussion.
Note that the natural units, $\hbar=c=1$, is used in this paper.

\section{THEORETICAL METHODS}
\label{sec:two}
In this section, we will briefly introduce the low-energy approximation model, which describes EH pair production in the vicinity of the Dirac point of graphene for an electric field.
This model is based on the assumption that the energy and momentum of electrons approximately satisfy the linear dispersion relation $E(\boldsymbol{p})=v_F|\boldsymbol{p}|+O[( \boldsymbol{p}/\boldsymbol{K}_{\pm})^2] $ \cite{Wallace1947}, where $v_F\approx 0.0036$ is the Fermi velocity \cite{Deacon2007}, $\boldsymbol{K}_{\pm}$ ($\left| \boldsymbol{K}_{\pm} \right|\approx 3361\mathrm{eV}$) is the position of the Dirac point in momentum space, $\boldsymbol{p}$ is the momentum of the particle with respect to $\boldsymbol{K}_{\pm}$.
It can be seen that the linear dispersion relation holds when $\boldsymbol{p}$ is much smaller than $\boldsymbol{K}_{\pm}$.
Thus, the maximum value of the momentum in our calculations is about $150\mathrm{eV}$.

Before obtaining a more accurate model, we neglected the effect of the magnetic field and assumed that the electric field acting on the graphene plane is space homogeneous.
By using the temporal gauge $A_{0}=0$, the vector potential can be written as $A_{\mu}(t)=(0, \boldsymbol{A}(t))$, where $\boldsymbol{A}(t)=(A_x(t), A_y(t))$.
The electric field is obtained by taking the derivative of the vector potential with respect to time, i.e., $\boldsymbol{E}(t)=-\dot{\boldsymbol{A}}(t)$.
In the presence of a time-dependent external electric field, the low energy excited state of graphene satisfies the massless Dirac equation \cite{Gusynin2007}
\begin{eqnarray}\label{eq:DiracEq}
i\partial _t\psi \left( t,\mathbf{p} \right) =v_F\boldsymbol{\sigma }\cdot \left[ \hat{\mathbf{p}}-e\boldsymbol{A}\left( t \right) \right] \psi \left( t,\mathbf{p} \right) ,
\end{eqnarray}
where $\boldsymbol{\sigma}$ is the Pauli matrix and $\hat{\mathbf{p}}$ is the momentum operator.

According to Eq. (\ref{eq:DiracEq}) and taking into account the condition of electrical neutrality $f^e(\boldsymbol{p}, t)=f^h(\boldsymbol{p}, t)=f(\boldsymbol{p}, t)$ (where $f^e(\boldsymbol{p}, t)$ and $f^h(\boldsymbol{p}, t)$ are the momentum distribution functions of electrons and holes, respectively), the kinetic equations describing the production and evolution of electrons (or holes) can be obtained \cite{Smolyansky2019,Aleksandrov2020,Gavrilov2020,Smolyansky2020}
\begin{eqnarray}\label{eq:f}
\dot{f}\left( \boldsymbol{p},t \right)&=&\frac{1}{2}\lambda \left( \boldsymbol{p},t \right) \int_{-\infty}^t{dt'\lambda \left( \boldsymbol{p},t' \right) \left[ 1-2f\left( \boldsymbol{p},t' \right) \right]}\nonumber\\
&&\times \cos\left[2\int_{t'}^t{dt''\varepsilon (\boldsymbol{p},t'')}\right],
\end{eqnarray}
where $\lambda(\boldsymbol{p}, t)=ev_{F}^{2}[E_x(t)q_y(t)-E_y(t)q_x(t)]/\varepsilon ^2(\boldsymbol{p},t)$, $\boldsymbol{q}(t)=\boldsymbol{p}-e\boldsymbol{A}(t)$ is the kinematic momentum, and $\varepsilon(\boldsymbol{p},t)=v_F\left|\boldsymbol{q}(t)\right|$ is the quasi-energy.

For the convenience of numerical calculation, two auxiliary variables $u(\boldsymbol{p},t)$ and $v(\boldsymbol{p},t)$ are introduced
\begin{eqnarray}\label{eq:uv}
\begin{array}{l}
u\left( \boldsymbol{p},t \right)= \int_{-\infty}^t{dt'{\lambda \left( \boldsymbol{p},t' \right) \left[ 1-2f\left( \boldsymbol{p},t' \right) \right]}}\\
\\
\quad \quad \quad \quad     \times  \cos\left[2\int_{t'}^t{dt''\varepsilon (\boldsymbol{p},t'')}\right],
\\
\\
v\left( \boldsymbol{p},t \right)= \int_{-\infty}^t{dt'{\lambda \left( \boldsymbol{p},t' \right) \left[ 1-2f\left( \boldsymbol{p},t' \right) \right]}}\\
\\
\quad \quad \quad \quad     \times  \sin\left[2\int_{t'}^t{dt''\varepsilon (\boldsymbol{p},t'')}\right],
\end{array}
\end{eqnarray}
then equation (\ref{eq:f}) can be equivalently transformed into the following first-order differential equations
\begin{eqnarray}\label{eq:fuv}
\begin{array}{l}
\dot{f}\left( {\boldsymbol{p},t} \right)= \frac{1}{2}\lambda\left( {\boldsymbol{p},t} \right)u\left( {\boldsymbol{p},t} \right),\\
\\
\dot{u}\left( {\boldsymbol{p},t} \right)= \lambda\left( {\boldsymbol{p},t} \right)\left[ {1 - 2f\left( {\boldsymbol{p},t} \right)} \right] - 2\varepsilon\left( {\boldsymbol{p},t} \right)v\left( {\boldsymbol{p},t} \right),\\
\\
\dot{v}\left( {\boldsymbol{p},t} \right)= 2\varepsilon\left( {\boldsymbol{p},t} \right)u\left( {\boldsymbol{p},t} \right).
\end{array}
\end{eqnarray}
In the initial state of the system, the external electric field is $0$, and the vector potential satisfies $\boldsymbol{A}(-\infty)=0$, so the momentum distribution function and the auxiliary functions satisfy the initial condition $f(\boldsymbol{p}, -\infty)=u(\boldsymbol{p}, -\infty)=v(\boldsymbol{p}, -\infty)=0$.

For an electric field that has only an $x$ component, $p_y$ can act as a mass term, so the Keldysh parameter can be defined similarly to the EP pair production case \cite{Keldysh1965,Brezin1970}: $\gamma=\left|p_y\right|\omega/eE_0$ , where $E_0$ and $\omega$ are the amplitude and frequency of the electric field, respectively.
When $\gamma\gg1$, the multi-photon process plays a dominant role. The momentum distribution function will have a local maximum value if the condition of multi-photon resonance absorption
\begin{eqnarray}\label{eq:MPAC}
2v_F\left| \boldsymbol{p} \right|=N\omega
\end{eqnarray}
is satisfied \cite{Kohlfurst2014,Avetissian2012,Mocken2010}, where $N$ is the number of absorbed photons.
When $\gamma\ll1$, the tunneling mechanism plays a dominant role.
Specifically, $\gamma=0$ for $p_y=0$. In this case, the distribution function has an analytical expression \cite{Fillion-Gourdeau2015}
\begin{eqnarray}\label{eq:fpx}
f\left( \boldsymbol{p},+\infty \right) \mid_{p_y=0}^{\,\,}=\mathrm{Rect}\left( \frac{p_x}{eA_x\left( +\infty \right)}-\frac{1}{2} \right),
\end{eqnarray}
where $\mathrm{Rect}$ is the rectangular function
\begin{eqnarray}\label{eq:rect}
\mathrm{Rect}\left( \mathrm{x} \right) =\begin{cases}
	1, \quad \left| x \right|\leqslant \frac{1}{2}\\
	0, \quad \left| x \right|>\frac{1}{2}\\
\end{cases}.
\end{eqnarray}
These equations indicate that the value of the momentum distribution function is $1$ when $p_x\in [eA_x(+\infty), 0]$.

The electric field we considered is a spatially homogeneous and time-dependent electric field, which consists of two elliptically polarized electric fields with a time delay
\begin{eqnarray}\label{eq:Eext}
\mathbf{E}\left( t \right)&=&E_{10}g_{\tau}(t) \left( \begin{array}{c}
	\cos \left( \omega t+\varphi_1  \right)\\
	\delta _1\sin \left( \omega t+\varphi_1 \right)\\
\end{array} \right)\nonumber\\
 &&+E_{20}g_{\tau}(t-T) \left(
\begin{array}{c}
	\cos \left[ \omega (t-T)+\varphi_2 \right]\\
	\delta _2\sin \left[ \omega (t-T)+\varphi_2 \right]\\
\end{array} \right) ,
\end{eqnarray}
where $g_{\tau}(t)=\exp[-t^2/(2\tau^2)] $, $\tau$ denotes the pulse duration, $E_{10,20}=E_{1,2}/\sqrt{1+\delta _{1,2}^{2}}$ are the amplitudes of electric fields, $\delta _{1,2}\in[-1, 1]$ are the field ellipticity, $\omega$ represents the field frequency, $T$ is the time delay between the two fields, and $\varphi_{1,2}$ are the carrier envelope phases. The value of $\varphi_{1,2}$ is set to $0$ if not specified.
Besides, a dimensionless quantity $\sigma =\omega \tau $ is introduced to characterize the number of oscillation cycles in a single electric field pulse.
In particular, when $E_2=0$, equation (\ref{eq:Eext}) represents a single elliptically polarized electric field.

\section{SINGLE ELLIPTICALLY POLARIZED FIELD}
\label{sec:three}

To analyze the result of EH pair production for two arbitrarily polarized electric fields with a time delay conveniently, we consider EH pair production dominated by multi-photon resonance absorption process for a single elliptically polarized electric field in this section. So the value of $E_2$ is set to $0$ in Eq. (\ref{eq:Eext}).

\begin{figure}[!ht]
\centering
\includegraphics[width=0.45\textwidth]{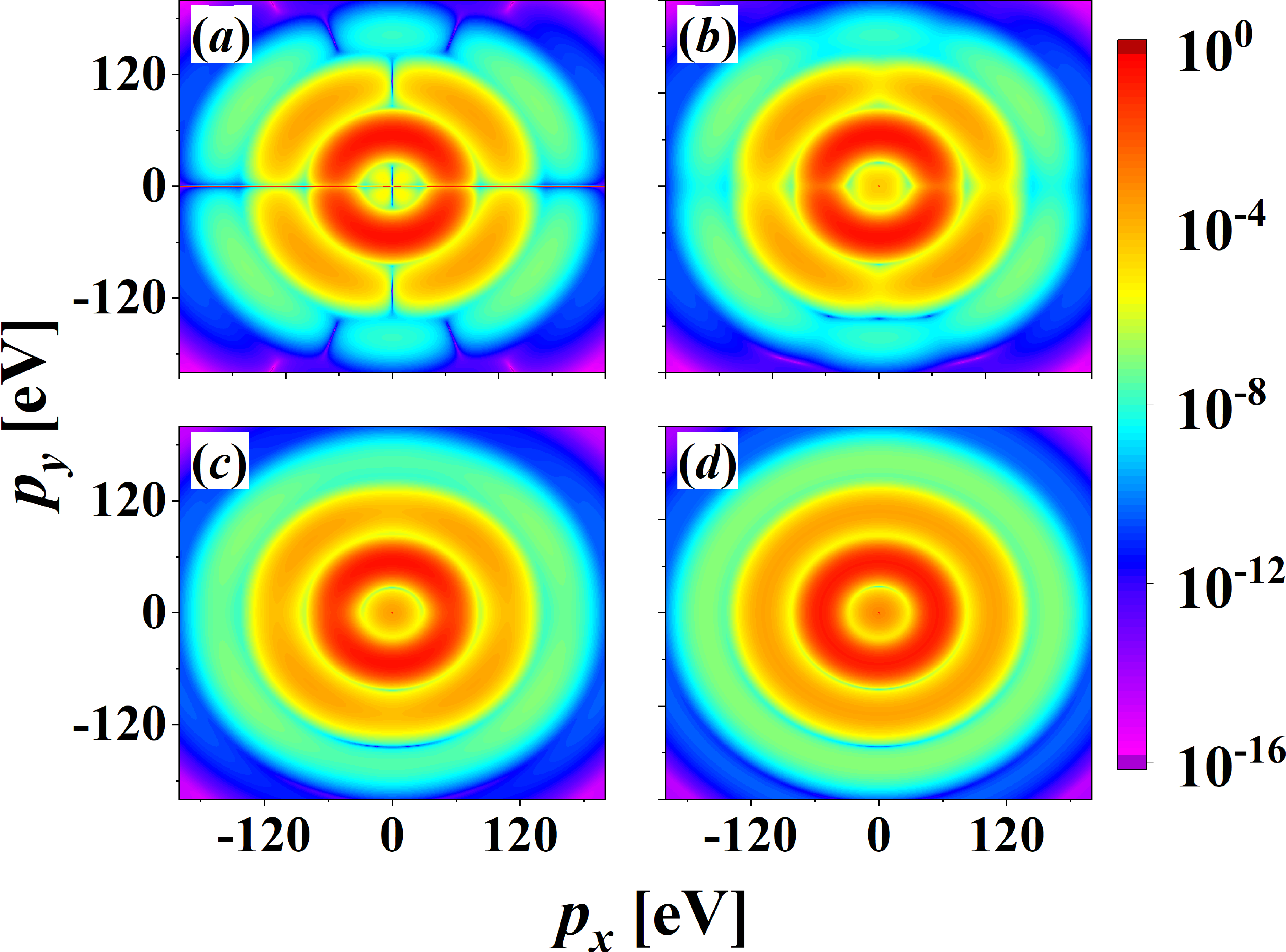}%
\caption{The momentum distribution of created EH pairs for a single elliptically polarized electric field ($E_2=0$) with the polarization values (a) $\delta_1=0.0$, (b) $\delta_1=0.1$, (c) $\delta_1=0.3$, and (d) $\delta_1=1.0$. Other electric field parameters are $E_{10}=\sqrt{2/(1+\delta _{1}^{2})}\times 10^7 \mathrm{V}/\mathrm{m}$, $\omega_1=622.0 \mathrm{THz}$, and $\tau=11.3 \mathrm{fs}$.
\label{fig:f1}}
\end{figure}

Figure \ref{fig:f1} shows the momentum distribution of created EH pairs for the polarization value $\delta_1=0$, $0.1$, $0.3$ and $1.0$, respectively. Other electric field parameters are $E_{10}=\sqrt{2/(1+\delta _{1}^{2})}\times 10^7\mathrm{V}/\mathrm{m}$, $\omega_1=622.0\mathrm{THz}$, and $\tau=11.3\mathrm{fs}$.
The ring structure on the momentum spectrum can be seen in the figure, which is formed due to $N$-photon absorption, where $N$ is the number of absorbed photons.
According to the resonance condition Eq. (\ref{eq:MPAC}), it can be estimated that the momentum values corresponding to the $1$-, $2$- and $3$-photon absorption are $56.15\mathrm{eV}$, $112.3\mathrm{eV}$ and $168.45\mathrm{eV}$, respectively.
This estimation is in accordance with the position of each ring in Fig. \ref{fig:f1}.
Furthermore, we can see that the value of the momentum distribution function will decrease several orders of magnitude for every one increase in the number of photons.
The EH pair production is dominated by $1$-photon absorption process.
However, for EP pair production by absorbing $1$ photon, the field frequency is required up to $\omega >2m\approx 7.76\times 10^8\mathrm{THz}$, according to the conservation of energy, $N\omega=2\sqrt{m^2+\boldsymbol{p}^2}$, where $m$ is the mass of the electron.

For a linearly polarized electric field ($\delta_1=0.0$), the Keldysh adiabatic parameter is much smaller than $1$ near $p_y=0.0$.
So the EH pair production near the line $p_y=0.0$ in momentum distribution is dominated by tunneling mechanism.
Furthermore, since the final value of the vector potential in Fig. \ref{fig:f1}(a) is $eA_x(+\infty)\approx3.0\times10^{-9}\mathrm{eV}$, according to Eq. (\ref{eq:fpx}) the momentum distribution function at $p_y=0.0$ reaches a saturation value in the interval $0.0\leqslant p_x\leqslant 3.0\times 10^{-9}\mathrm{eV}$ and is $0.0$ at other positions. Therefore, the momentum distribution function in Fig. \ref{fig:f1}(a) is cut along $p_y=0.0$ except the very small saturation interval.
Moreover, other node structures (the momentum distribution function tends to $0$) can also be seen on the multiphoton absorption ring, i.e., at $p_x=0$ for the $2$-photon absorption and $p_x\approx\pm56\mathrm{eV}$ for the $3$-photon absorption.
This is very similar to the EP pair production in a vacuum for a strong external electric field.
According to Ref. \cite{Marinov1977}, the factor in the momentum distribution function for the EH pair production can be similarly written as $[1+(-1)^{N+1}\cos(4v_F|\boldsymbol{p}|/\omega\cdot\mathrm{arc}\tan(\gamma))]$. The node on the $3$-photon absorption ring calculated by the above factor is located at $p_x\approx 57.8\mathrm{eV}$. One can see that it is in good agreement with the result in Fig. \ref{fig:f1}(a).
In addition, from Figs. \ref{fig:f1}(b)-(d), it shows that the node structures on multiphoton absorption rings quickly disappear with the increase of the polarization value.
This is because, as the polarization value increases, the electric field gradually becomes isotropic.

\section{TWO ARBITRARILY POLARIZED FIELDS WITH A TIME DELAY}
\label{sec:four}
To further illustrate the similarities and differences between the EH pair production in graphene and the EP pair production in a vacuum for a strong electric field, we will investigate some special structures in the momentum distribution of created EH pairs using two arbitrarily polarized electric fields with a time delay in this section.
For convenience, in the following, we refer to the combination of a left-handed circularly polarized electric field ($\delta _1=1$) and a left-handed one ($\delta _2=1$) with a time delay as the LLCP electric field, and the combination of a left-handed circularly polarized electric field ($\delta _1=1$) and a right-handed one ($\delta _2=-1$) with a time delay as the LRCP electric field.

\subsection{Two linearly polarized electric fields}\label{A}

\begin{figure}[!ht]
\centering
\includegraphics[width=0.45\textwidth]{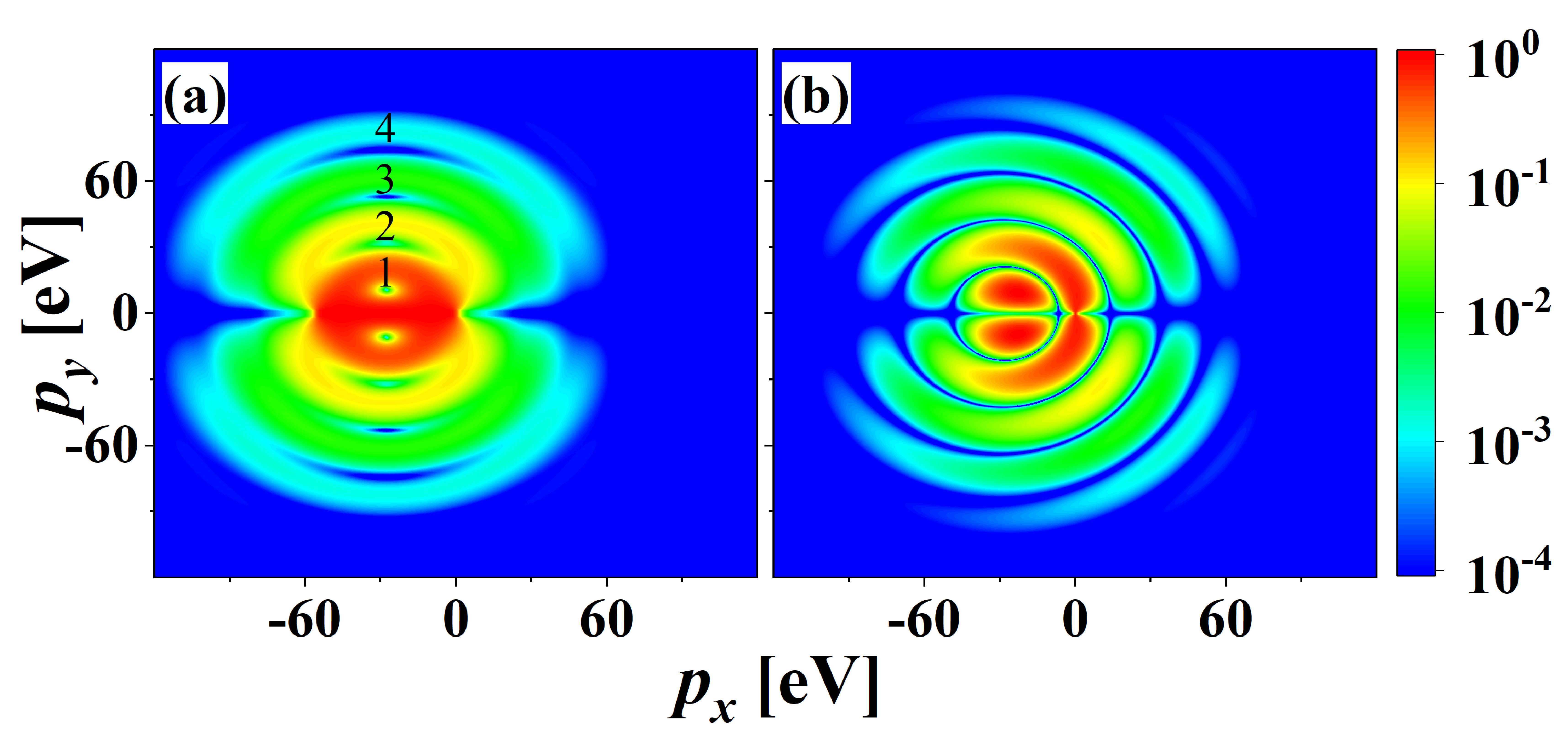}%
\caption{Momentum distribution of created EH pairs for two linearly polarized electric fields ($\delta _1=\delta_2=0$) with a time delay. (a) and (b) correspond to two co-directional electric fields, $E_{10}=E_{20}=1.5\times10^7\mathrm{V}/\mathrm{m}$, and two counter-directional electric fields, $E_{10}=-E_{20}=1.5\times10^7\mathrm{V}/\mathrm{m}$, respectively. Other field parameters are $\omega=0$, $\tau=2.25\mathrm{fs}$, and $T=12\tau$.
\label{fig:fexp}}
\end{figure}

The momentum distributions of created EH pairs for two co-directional ($E_{10}=E_{20}$) and counter-directional ($E_{10}=-E_{20}$) linearly polarized electric fields ($\delta_1=\delta_2=0$) with a time delay are shown in Fig. \ref{fig:fexp}(a) and Fig. \ref{fig:fexp}(b), respectively.
The other electric field parameters are $E_{10}=|E_{20}|=1.5\times10^7\mathrm{V}/\mathrm{m}$, $\omega=0$, $\tau=2.25\mathrm{fs}$, and $T=12\tau$.

\begin{figure}[!ht]
\centering
\includegraphics[width=0.45\textwidth]{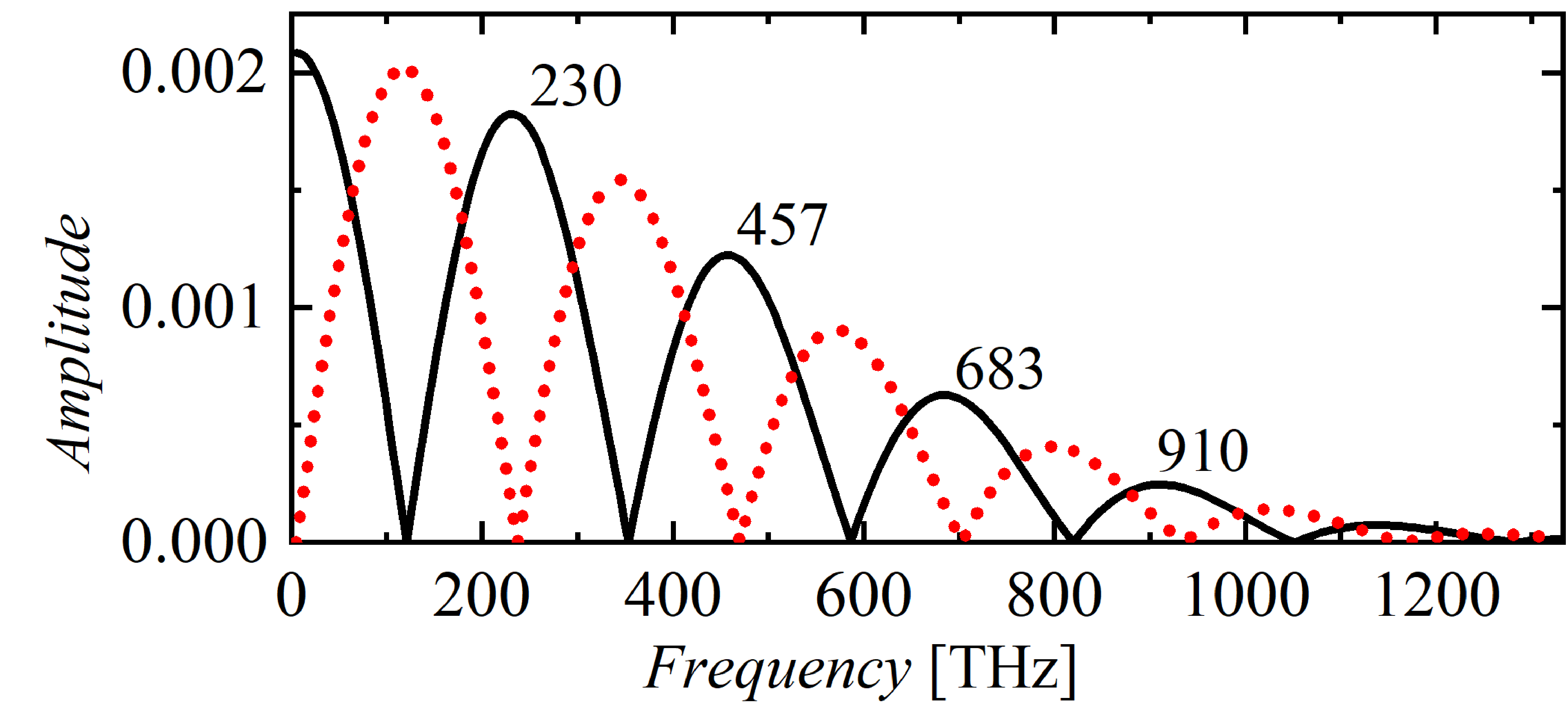}%
\caption{The Fourier transform of the electric field. The black solid line and the red dotted line correspond to the electric field parameters in Figs. \ref{fig:fexp}(a) and \ref{fig:fexp}(b), respectively.
\label{fig:frequency}}
\end{figure}

From Fig. \ref{fig:fexp}(a), we can see that there are two main differences between EH and EP pair production. 
Firstly, the momentum distribution of created EH pairs is not completely separated into two parts in the $x$-direction. 
It is known that the momentum distribution of created EP pairs for two co-directional electric fields forms two separate parts in the $x$-direction due to the fact that the second electric field will accelerate the particles produced by the first electric field.
However, for EH pair production, since the energy gap in graphene is $0$, a lot of EH pairs can always be produced by tunneling mechanism in the vicinity of the Dirac line, see the line segment from $(-55\mathrm{eV}, 0)$ to $(0, 0)$ in the $p_x$-$p_y$ plane.
Secondly, the momentum distribution of created EH pairs shows obvious ring patterns in the $y$-direction, which is significantly different from that in EP pair production \cite{Akkermans2011}.
Four rings are labeled in Fig. \ref{fig:fexp}(a) and the values of $p_y$ corresponding to them are $20.1\mathrm{eV}$, $40.9\mathrm{eV}$, $61.4\mathrm{eV}$, and $82.2\mathrm{eV}$, respectively.
In fact, the rings are caused by $1$-photon absorption process.
To explain this, we show the Fourier transform of the electric field used in Fig. \ref{fig:fexp}(a) in Fig. \ref{fig:frequency}, see the black solid line, where the positions of the four peaks are about $230\mathrm{THz}$, $457\mathrm{THz}$, $683\mathrm{THz}$, and $910\mathrm{THz}$, respectively.
According to Eq. (\ref{eq:MPAC}), the momenta corresponding to absorbing one photon with the above four different frequencies are about $21.0\mathrm{eV}$, $41.8\mathrm{eV}$, $62.4\mathrm{eV}$, and $83.2\mathrm{eV}$, respectively.
One can see that this result is well agree with the numerical one in Fig. \ref{fig:fexp}(a).

The momentum distribution of created EH pairs for two counter-directional linearly polarized electric fields is shown in Fig. \ref{fig:fexp}(b).
Firstly, we can see that the distribution function is $0$ at $p_y=0$, because the final vector potential is $0$.
Secondly, similar to the case of Fig. \ref{fig:fexp}(a), the momentum distribution has obvious $1$-photon absorption rings, which is caused by absorbing one photon with the frequencies corresponding to the peaks of the red dotted line in Fig. \ref{fig:frequency}. Note that the red dotted line is the Fourier transform of the electric field used in Fig. \ref{fig:fexp}(b).
Finally, the momentum distribution also has interference in the $x$ direction, which is a time-domain interference similar to the EP pair production \cite{Akkermans2011}.

\subsection{The LLCP electric field}\label{B}
\begin{figure}[!ht]
\centering
\includegraphics[width=0.23\textwidth]{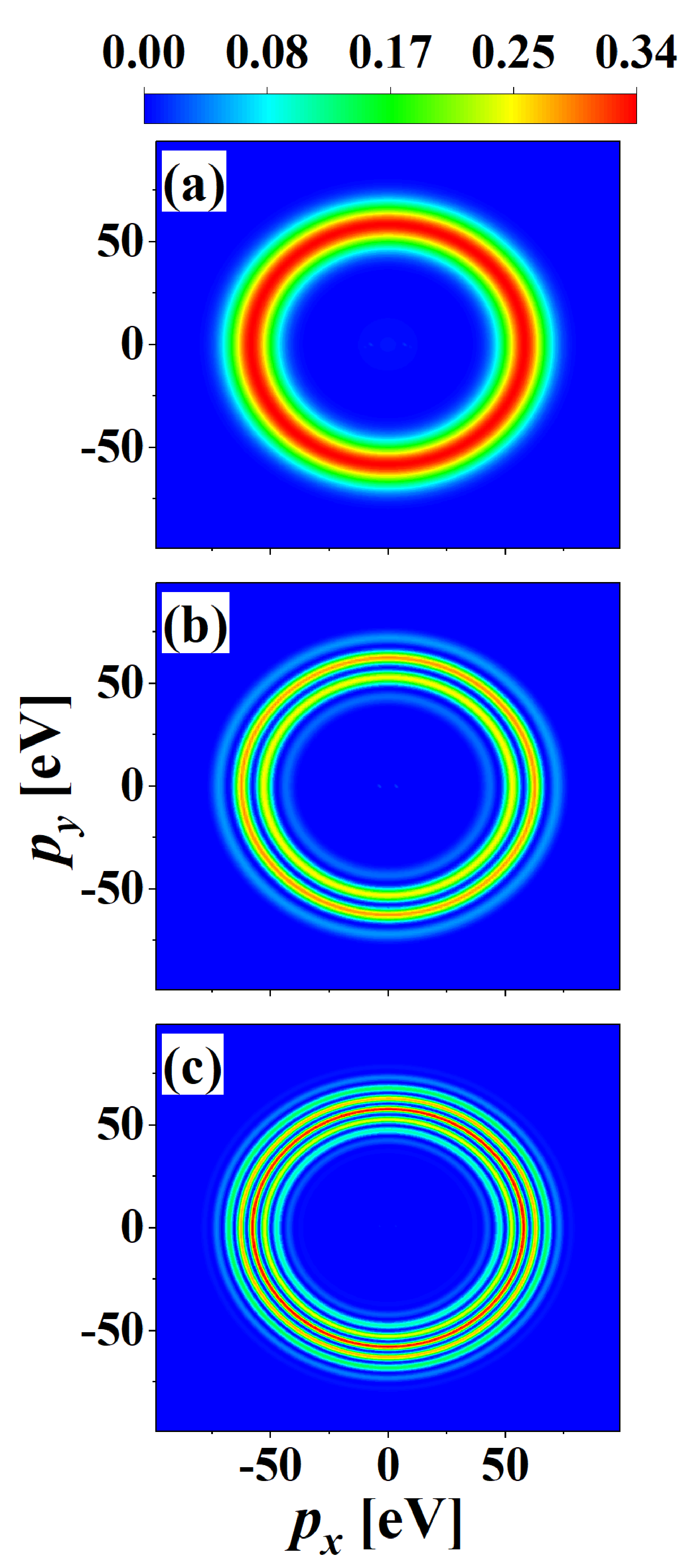}%
\includegraphics[width=0.23\textwidth]{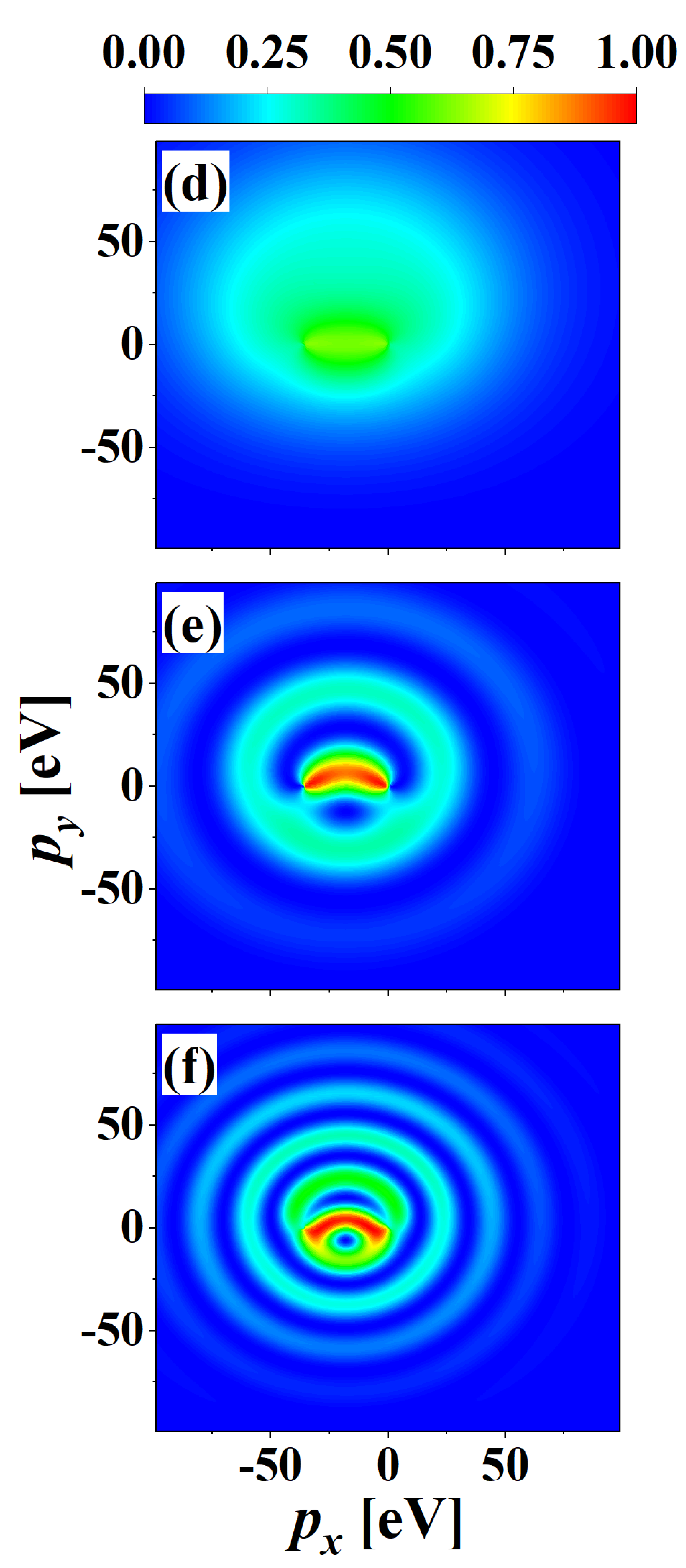}%
\caption{Momentum distribution of created EH pairs for the LLCP electric field with a time delay $T=0$ (first row), $6\tau$ (middle row), and $12\tau$ (last row). The electric field amplitudes and pulse durations are $E_{10}=E_{20}=1.0\times10^7\mathrm{V}/\mathrm{m}$, $\tau=9.0\mathrm{fs}$ for the first column, and $E_{10}=E_{20}=3.0\times10^7\mathrm{V}/\mathrm{m}$, $\tau=2.25\mathrm{fs}$ for the second column. The field frequency is $\omega=666.5\mathrm{THz}$ for all cases.
\label{fig:f2}}
\end{figure}

Figures \ref{fig:f2}(a)-(c) show the momentum distributions of created EH pairs for multi-cycle LLCP electric fields with a time delay (a) $T=0$, (b) $T=6\tau$, and (c) $T=12\tau$.
The other field parameters are $E_{10}=E_{20}=1.0\times10^7\mathrm{V}/\mathrm{m}$, $\tau=9.0\mathrm{fs}$, and $\omega=666.5\mathrm{THz}$.
In this case, the number of oscillation cycles in each electric field pulse is $\sigma=7$ and the final vector potential is $\boldsymbol{A}(+\infty)\approx0$.
In Fig. \ref{fig:f2}(a), the momentum distribution has a ring structure with a radius of $58.6\mathrm{eV}$, which is caused by $1$-photon absorption process. The radius of the ring can also be well predicted by Eq. (\ref{eq:MPAC}).
This result is similar to the situation discussed in Fig. \ref{fig:f1}(d).
Because the superposition of two left-handed electric fields is still a left-handed circularly polarized field for $T=0$.
But the situation is different when the time delay $T=6\tau, 12\tau$.
In Figs. \ref{fig:f2}(b) and (c), many concentric rings appear in the region of $1$-photon absorption ring.
As the time delay $T$ increases, the number of rings will increase, and each ring will become thinner.
These concentric rings result from the interference between the amplitudes of EH pairs produced in two left-handed electric fields, which will be explained in detail below.

In Figs. \ref{fig:f2}(d)-(f), the momentum distributions of created EH pairs for few-cycle LLCP electric fields with a time delay $T=0,\, 6\tau,\, 12\tau$ are shown.
The other field parameters are $E_{10}=E_{20}=3.0\times10^7\mathrm{V}/\mathrm{m}$, $\tau=2.25\mathrm{fs}$, and $\omega=666.5\mathrm{THz}$, respectively.
The number of oscillation cycles in each electric field pulse is $\sigma=1.5$.
In this case, each left-handed electric field has a major peak in the $x$-direction and two antisymmetric peaks in the $y$-direction.
Therefore, the final vector potential in the $x$-direction is about $-118.86\mathrm{eV}$, while it is $0$ in the $y$-direction.
This will lead to a translation of the momentum distribution in the $x$-direction.
As can be seen from these figures, the symmetrical axis of momentum distributions moves from $p_x=0$ to $-18.0\mathrm{eV}$.
Furthermore, since the number of oscillation cycles in each pulse is very small, the multi-photon absorption process is not obvious.
This can be seen from Fig. \ref{fig:f2}(d), where no multi-photon absorption rings are present on the momentum distribution.
From Fig. \ref{fig:f2}(d), we can also see that the maximum value of the momentum distribution at $p_y=0$ is about $0.64$ rather than $1$.
This is because the EH pairs are mainly produced at the major peak of the electric field (near $t=0$), and then accelerated along the positive $y$-direction by the peak of the electric field in $y$-direction at $t>0$.
In Figs. \ref{fig:f2}(e) and (f), where the time delays are $6\tau$ and $12\tau$, respectively, some interference rings appear on the momentum distribution, which are different from Figs. \ref{fig:f2}(b) and (c). These rings have a wider distribution and their positions are not in the region of the $1$-photon absorption ring.
Furthermore, the smaller the rings, the larger the value of the distribution functions.
In the following, we will discuss the reasons for the appearance of the rings in Fig. \ref{fig:f2} and give a theoretical formula that can determine the positions of the rings.

According to previous research on semiclassical approximation \cite{Dumlu2011,Akkermans2011,Li2014xga}, the particles produced in the LLCP electric field is dominated by the amplitudes of pair production near the turning points $t_0$ and $t_T$.
Here the turning points are defined as the solutions of the equation $\varepsilon(\boldsymbol{p}, t)=0$ for a given value of momentum $\boldsymbol{p}$.
The formation of concentric rings shown in Fig. \ref{fig:f2} can be understood as an interference between these two pairs of turning points.
Similar to EP pair production studied in Refs. \cite{Li2017qwd,Li2018hzi,Hu2023pmz}, for a given momentum $\boldsymbol{p}$, the amplitude of EH pair production for the first electric field is $A_1=\exp[-iK(\boldsymbol{p}, t_0)]$, and for the second field is $A_2=\exp[-iK(\boldsymbol{p}, t_T)]$, where $K(\boldsymbol{p}, t)=2\int_{-\infty}^t{dt^{\prime}\varepsilon (\boldsymbol{p}, t^{\prime})}$.
Thus the momentum distribution can be calculated by
\begin{eqnarray}\label{eq:fAA}
f\left( \boldsymbol{p} \right)&=&\left| A_1+A_2 \right|^2.
\end{eqnarray}
For a large $\sigma$, the electric field and vector potential at both $t=0$ and $t=T$ have completely identical profiles, so the amplitude of pair production corresponding to the second left-handed electric field for a larger $T$ can be rewritten as
\begin{eqnarray}\label{eq:A2}
A_2\approx\exp[i\zeta(\boldsymbol{p})]A_1,
\end{eqnarray}
where $\zeta(\boldsymbol{p})=\mathrm{Re}[K(\boldsymbol{p}, t_T)-K(\boldsymbol{p}, t_0 )]$ is the accumulation of phases between the two fields.
Then the momentum distribution becomes
\begin{eqnarray}\label{eq:fq}
f\left( \boldsymbol{p} \right)&=&\left| A_1+A_2 \right|^2\nonumber\\
&\approx&\left| A_1+\exp \left[ i\zeta \left( \boldsymbol{p} \right) \right] A_1 \right|^2\nonumber\\
&=&2\left( 1+\cos \left[ \zeta \left( \boldsymbol{p} \right) \right] \right) \exp \left( 2\mathrm{Im}\left[ K\left( \boldsymbol{p},t_0 \right) \right] \right).
\end{eqnarray}
For a large time delay $T$, $\zeta(\boldsymbol{p})$ can be approximated as $\zeta(\boldsymbol{p})\approx2v_{F}\left| \boldsymbol{q}\left( T/2 \right) \right|T$ because the electric field and vector potential between $t=0$ and $t=T$ are constants.
Therefore, the concentric rings appear at $\cos[\zeta(\boldsymbol{p})]=1$, i.e.,
\begin{equation}\label{eq:pring}
\sqrt{[ p_x-eA_x\left( T/2 \right)] ^2+[ p_y-eA_y\left( T/2 \right) ]^2}=\frac{k\pi}{v_FT},
\end{equation}
where $k$ is an integer.

\begin{figure}[!ht]
\centering
\includegraphics[width=0.45\textwidth]{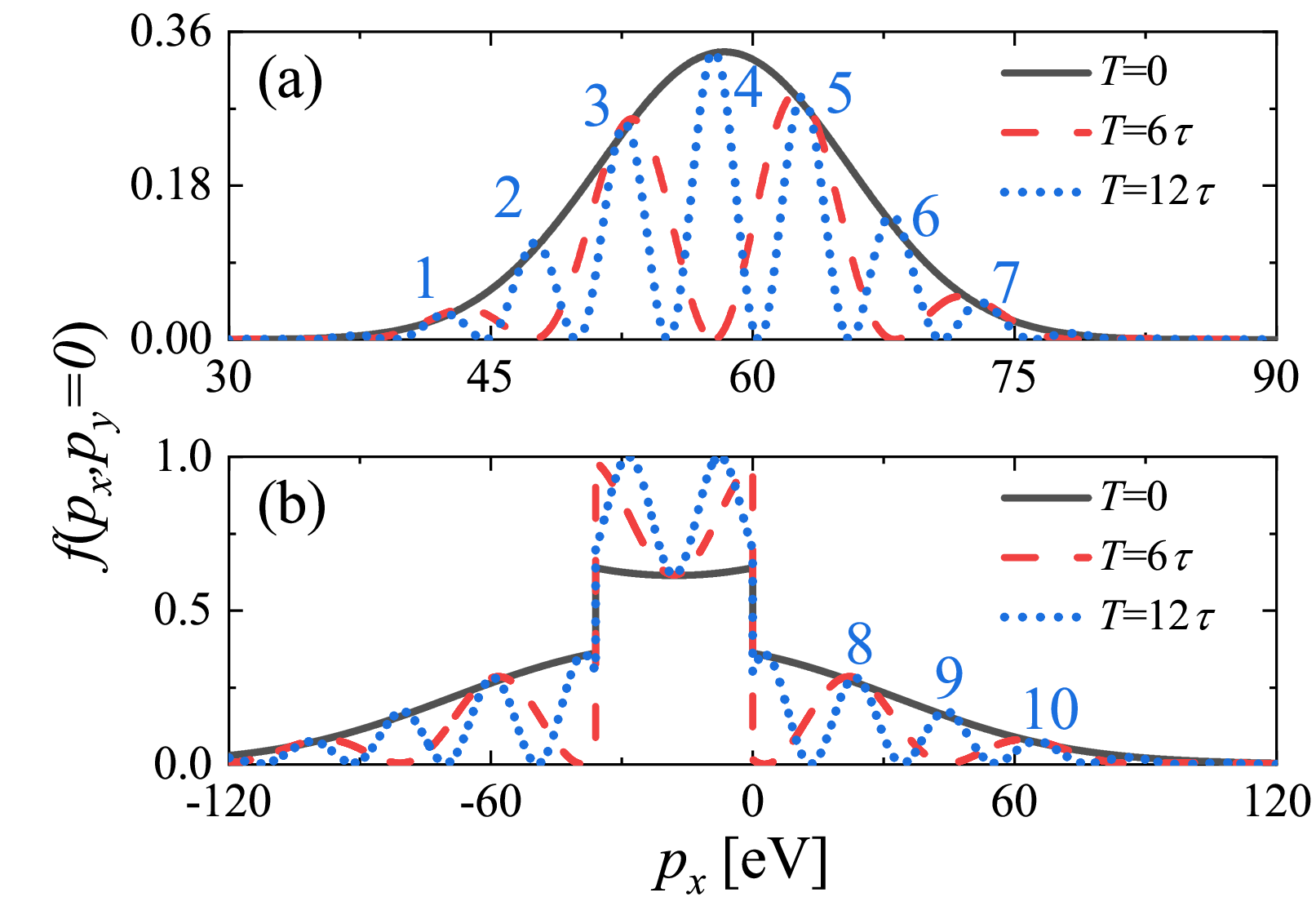}%
\caption{Momentum distribution at $p_y=0$. The black solid line, red dashed line, and blue dotted line correspond to $T=0$, $6\tau$, and $12\tau$, respectively. The peaks on the momentum distribution for $T = 12$ are marked with numbers from $1$ to $10$. The electric field parameters are the same as in Figs. \ref{fig:f2}(a)-(c) for (a) and Figs. \ref{fig:f2}(d)-(f) for (b).
\label{fig:f2py0}}
\end{figure}

In Fig. \ref{fig:f2py0}, the momentum distribution function at $p_y=0$ is given. The electric field parameters are the same as in Fig. \ref{fig:f2}.
In the figure, the peaks on the momentum distribution for $T=12\tau$ is marked with the number $i$ ($1$ to $7$ for (a) and $8$ to $10$ for (b)).
To check the degree of agreement between Eq. (\ref{eq:pring}) and the numerical results, we show in Table \ref{tab:comparison} the momentum value, $p_{x}^{\mathrm{N}}$, obtained by numerical computation and the momentum value, $p_{x}^{\mathrm{T}}$, estimated by Eq. (\ref{eq:pring}) at each peak. Furthermore, the degree of their error, $\Delta p_x=\left|p_{x}^{\mathrm{T}}-p_{x}^{\mathrm{N}}\right|$, is also calculated. From the table, we can see that the positions of rings can be well determined by Eq. (\ref{eq:pring}).

\begin{table*}[!ht]\suppressfloats
\renewcommand{\arraystretch}{1.5}
\setlength{\tabcolsep}{11pt}
\begin{center}
   \caption{\label{tab:comparison} Comparison between the results $p_{x}^{\mathrm{T}}$ computed by Eq. (\ref{eq:pring}) and the numerical results $p_{x}^{\mathrm{N}}$ for the $10$ peaks in Fig. \ref{fig:f2py0}.
   The values of $k$ for $i = 1$, $2$,..., $7$ are $8$, $9$,..., $14$, and for $i = 8$, $9$, $10$ are $2$, $3$, $4$.}
   \begin{tabular}{ccccccccccc} \hline \hline
    $i$   & $1$ & $2$ & $3$ & $4$ & $5$ & $6$ & $7$ & $8$ & $9$ & $10$ \\ \hline
    $p_{x}^{\mathrm{N}}\,\,[\mathrm{eV}]$ & $42.45$ & $47.57$ & $52.68$ & $57.76$ & $62.85$ & $67.95$ & $73.08$ & $23.46$ & $44.22$ & $65.01$ \\ 
    $p_{x}^{\mathrm{T}}\,\,[\mathrm{eV}]$ & $42.00$ & $47.25$ & $52.50$ & $57.75$ & $63.00$ & $68.25$ & $73.50$ & $24.00$ & $45.00$ & $66.00$ \\ 
    $\Delta p_x\,\,[\mathrm{eV}]$ & $0.45$ & $0.32$ & $0.18$ & $0.01$ & $0.15$ & $0.30$ & $0.42$ & $0.54$ & $0.78$ & $0.99$ \\  \hline \hline
   \end{tabular}
   \end{center}
\end{table*}

\subsection{The LRCP electric field}\label{C}
\begin{figure}[!ht]
\centering
\includegraphics[width=0.23\textwidth]{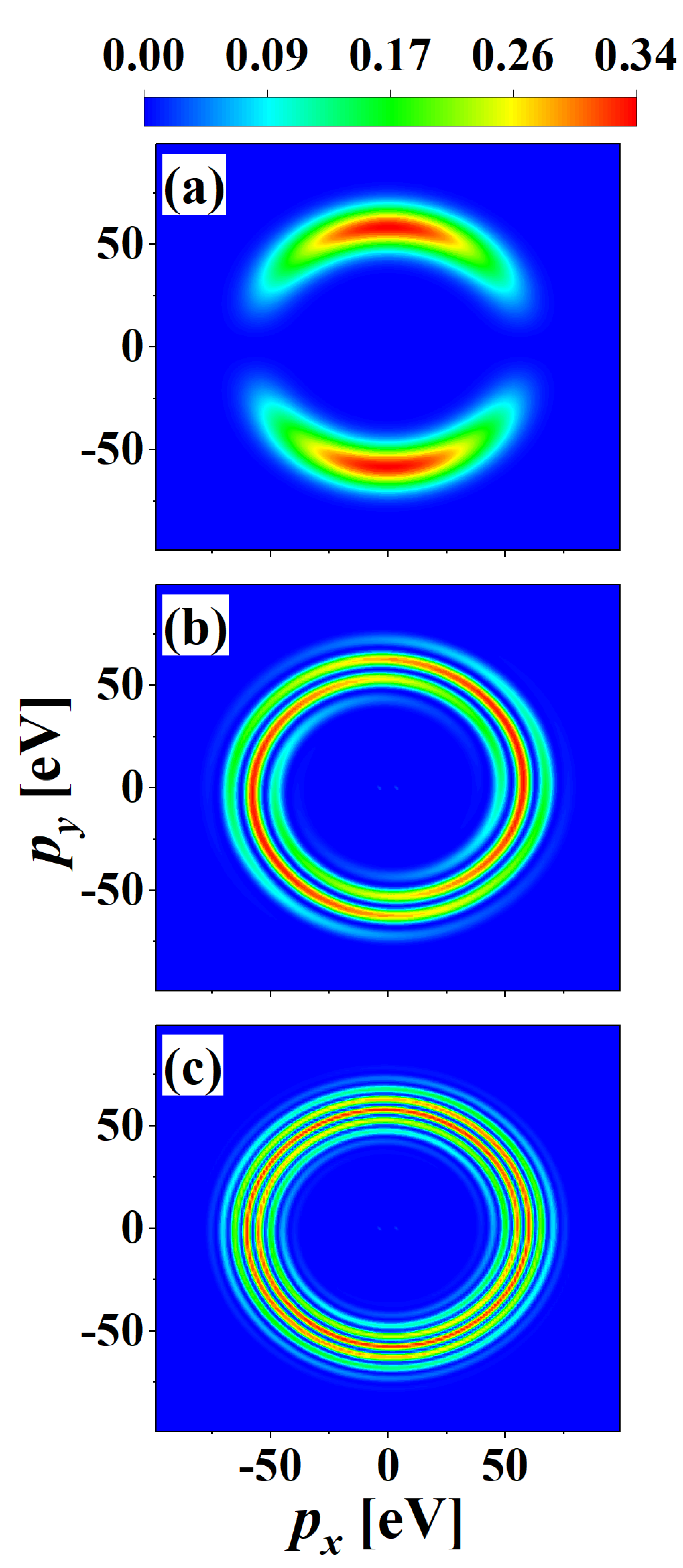}%
\includegraphics[width=0.23\textwidth]{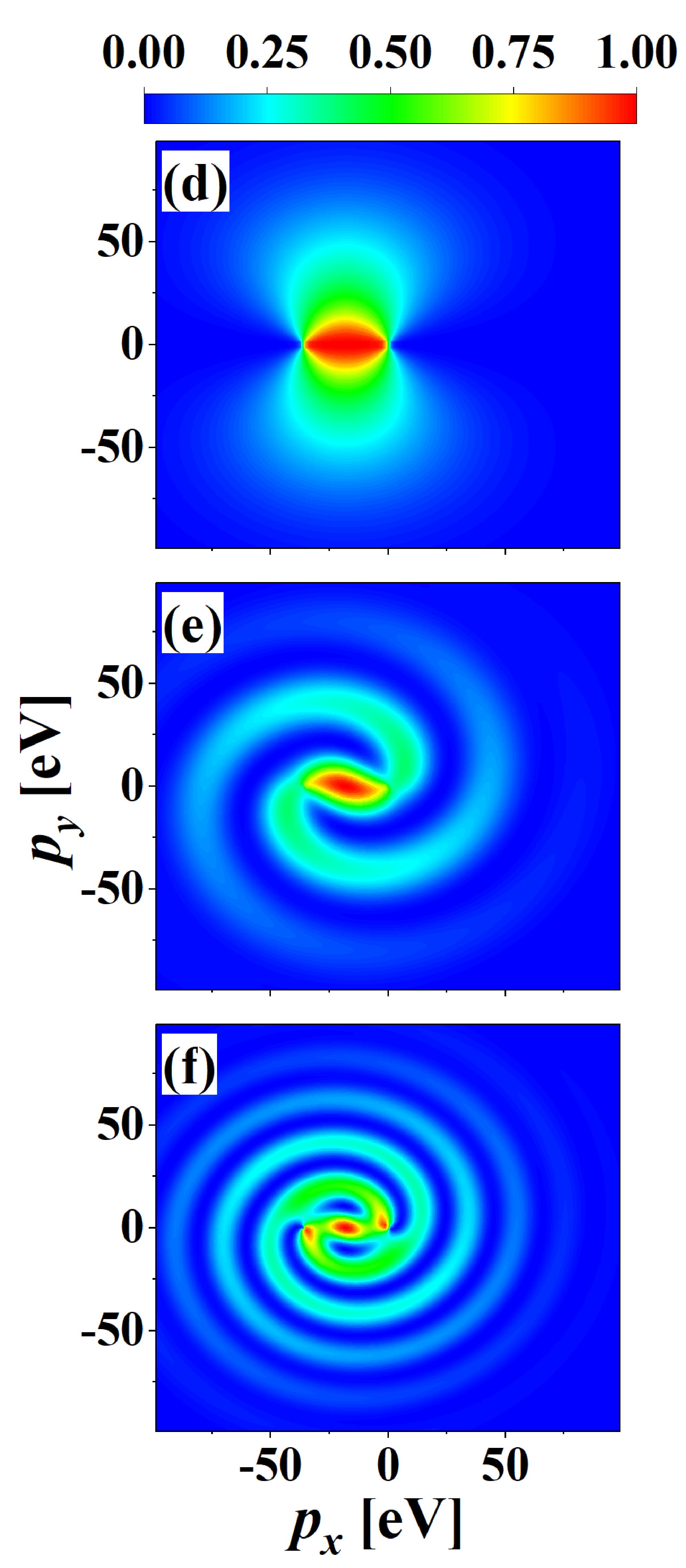}%
\caption{Momentum distribution of EH pairs created in graphene for a LRCP electric field. The field parameters are the same as in Fig. \ref{fig:f2}.
\label{fig:f3}}
\end{figure}

The momentum distribution of created EH pairs for a LRCP electric field is calculated and shown in Fig. \ref{fig:f3}.
The electric field parameters are the same as in Fig. \ref{fig:f2}, except for $\delta _1=-\delta _2=1$.
For the case in Fig. \ref{fig:f3}(a), where $T=0$, the superposition of two circularly polarized fields with opposite handedness forms a linearly polarized field. This is similar to the case in Fig. \ref{fig:f1}(a), i.e., multi-photon absorption rings appear on the momentum distribution.
When the time delays $T$ are $6\tau$ and $12\tau$, the momentum distributions have spiral structures in the region of $1$-photon absorption ring, see Figs. \ref{fig:f3}(b) and (c).
To explain the formation of the spirals, the theory in \cite{Li2017qwd} can still be referenced.
However, in order to suit the physical model of graphene, the polar coordinates $(p,\varphi)$ are used, where $p$ is the magnitude of momentum and $\varphi$ is the azimuthal angle.
The amplitude of created EH pairs for the first electric field is $A_1\approx \exp(il\delta _1\varphi-i\delta_1\pi/2)A_0(p,\varphi)$, where $l$ is the number of photons absorbed in EH pair production, and $\exp(-i\delta_1\pi/2)$ indicates the phase difference of the $y$-component of the electric field with respect to the $x$-component.
Similar to what we considered previously in the LLCP scenario, the amplitude of created EH pairs for the second field is $A_2\approx\exp(i\zeta(p,\varphi))\exp(il\delta_2\varphi-i\delta_2\pi/2)A_0(p,\varphi)$, where $\exp(i\zeta(p,\varphi))$ is the accumulation of phases between the two fields.
Thus, the momentum distribution is
\begin{eqnarray}\label{eq:fqphi}
f(p,\varphi)&\approx&\{ 1+\cos \left[ \zeta \left( p,\varphi \right) +\left( \delta _2-\delta _1 \right)( l\varphi -\pi/2) \right] \}\nonumber\\
&&\times A_0\left( p,\varphi \right).
\end{eqnarray}
When $T\gg \tau $, $\zeta(\boldsymbol{p})\approx2v_{F}qT$, where $q=\left| \boldsymbol{q}\left( T/2 \right) \right|$ is the magnitude of the kinetic momentum at $t=T/2$. For an electric field with a large $\sigma$, we have $p\approx q$.
From Eq. (\ref{eq:fqphi}), we obtain the equation of spirals on the momentum distribution,
\begin{eqnarray}\label{eq:vortex}
q_{k'}\left( \varphi \right) = \frac{2k'\pi -\left( \delta _2-\delta _1 \right) \left( l\varphi -\pi /2 \right)}{2v_FT},
\end{eqnarray}
where $k'$ is an integer that ensures that the right hand side of the equation is a positive real number.
Equation (\ref{eq:vortex}) represents a polar coordinate equation for an Archimedean spiral.
It gives the momentum distribution with clockwise spirals for $\delta _1=-\delta _2=1$ and counterclockwise spirals for $\delta _1=-\delta _2=-1$.
Since in Figs. \ref{fig:f3}(b) and (c), the EH pairs are mainly produced by the $1$-photon resonance absorption process, the spirals distribute in the region of the $1$-photon absorption ring.
As the time delay $T$ increases, the distance $\pi/T$ between the spirals becomes smaller, and the spirals become thinner and longer.
Meanwhile, the difference between the momentum distributions for the LLCP and LRCP field also becomes smaller. For instance, the momentum distribution in Fig. \ref{fig:f3}(c) and Fig. \ref{fig:f2}(c) are very similar when $T=12\tau$.
Moreover, since Eq. (\ref{eq:vortex}) is formally equal to Eq. (\ref{eq:pring}) minus $(\delta _2-\delta _1)(l\varphi -\pi /2)/{2v_FT}$, we find that when $\varphi=0$ the local maximum (local minimum) value of the momentum distribution for the LRCP appears at the local minimum (local maximum) value for the LLCP, while the positions of the maxima (minima) of the momentum distributions for the two cases are the same when $\varphi=\pi/2$.

\begin{figure}[!ht]
\centering
\includegraphics[width=0.45\textwidth]{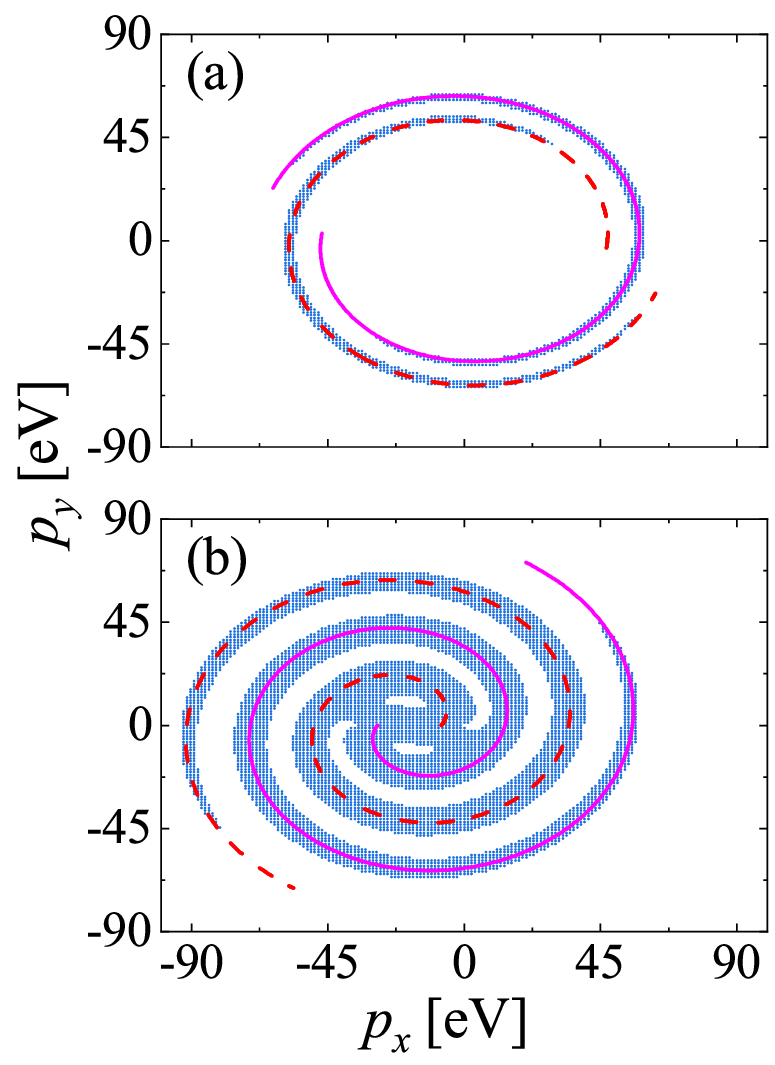}%
\caption{Comparison between the spirals given by Eq. (\ref{eq:pring}) and by numerical calculation. (a) and (b) correspond to the spiral patterns in Figs. \ref{fig:f3}(b) and (f), respectively. The blue dots indicate the result of the numerical calculation. The red dashed and magenta solid lines are the results given by Eq. (\ref{eq:pring}).}
\label{fig:spiral}
\end{figure}

The momentum distribution for the number of oscillation cycles $\sigma=1.5$ is shown in Figs. \ref{fig:f3}(d)-(f).
In Fig. \ref{fig:f3}(d), we can see that the EH pairs is mainly produced by tunneling mechanism, and the interval of momentum values of the distribution function approaching $1$ is compatible with that given by Eq. (\ref{eq:fAA}).
In Figs. \ref{fig:f3}(e) and (f), they show visible spiral structures which are significantly different from the results in EP pair production (see Fig. 7 in Ref. \cite{Hu2023pmz}).
For EP pair production, particle pairs are produced primarily at the main peak of the electric field because the EP pair production is exponentially suppressed for the electric field strength.
However, the results show that the requirement of the electric field strength to form spiral structures for EH pair production is weaker because EP pairs are massless.
Moreover, the distribution of the spirals shown in Figs. \ref{fig:f3}(e) and (f) is so extensive that it goes beyond the region of the $1$-photon absorption ring.
Nevertheless, the appearance of the spirals can still be determined by Eq. (\ref{eq:pring}) with $l=1$, see Fig. \ref{fig:spiral}(b). This is because although the $1$-photon absorption process is not obvious for the few-cycle LRCP field, it still play a role in EH pair production.

In Figs. \ref{fig:spiral}(a) and (b), we show the data points that the value of the momentum distribution is larger than $0.2$ for Fig. \ref{fig:f3}(b) and $0.1$ for Fig. \ref{fig:f3}(f).
The theoretical values of Eq. (\ref{eq:vortex}) are also plotted in Fig. \ref{fig:spiral} using the magenta solid and red dashed lines.
The red dashed line in Fig. \ref{fig:spiral}(a) (or \ref{fig:spiral}(b)) is obtained by choosing $k'=1$ and varying the azimuthal angle $\varphi$ from $12.5$ to $18.5$ (or from $0$ to $10.5$), while the magenta solid line is obtained by choosing $k'=0$ and varying the azimuthal angle $\varphi$ from $12.5+\pi$ to $18.5+\pi$ (or from $\pi$ to $10.5+\pi$).
From the figure, one can see that the theoretical results of Eq. (\ref{eq:pring}) are well agree with the numerical results.
Furthermore, since the effect of the tunneling mechanism, the spirals starts from an azimuthal angle $\varphi=0$ in Fig. \ref{fig:spiral}(b).

\subsection{Two elliptically polarized electric fields}\label{D}
\begin{figure}[!ht]
\centering
\includegraphics[width=0.24\textwidth]{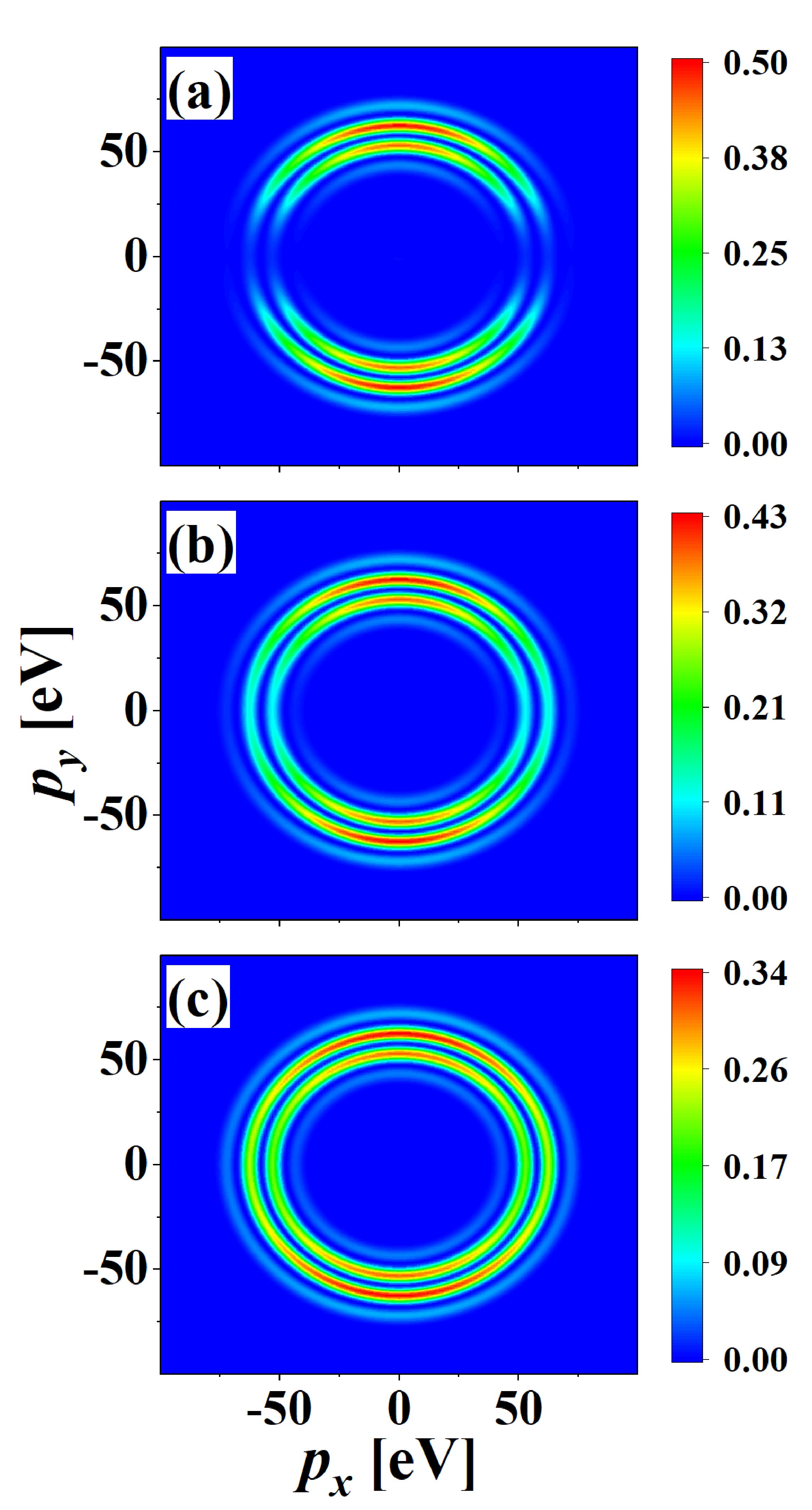}%
\includegraphics[width=0.24\textwidth]{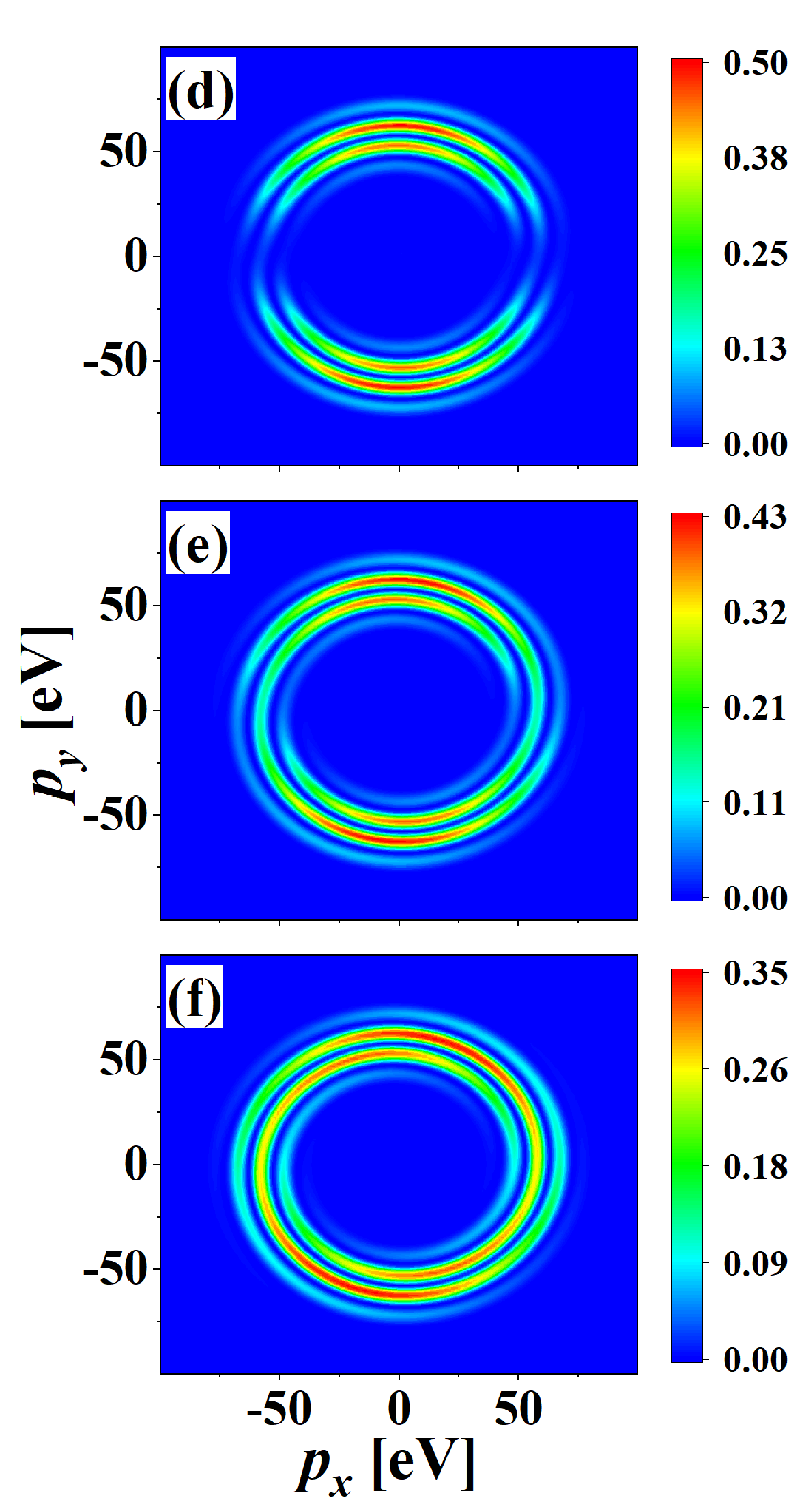}%
\caption{Momentum distribution of created EH pairs for two elliptically polarized electric fields with a time delay. For (a), (b), and (c), the ellipticity $\delta_1=\delta_2=0.2$, $0.5$, and $0.8$, respectively. For (d), (e), and (f), the ellipticity $\delta_1=-\delta_2=0.2$, $0.5$, and $0.8$, respectively. The other electric field parameters are $E_1=E_2=1.0\times10^7\mathrm{V}/\mathrm{m}$, $\omega=666.5\mathrm{THz}$, $\tau=9.0\mathrm{fs}$, and $T=6\tau$.
\label{fig:pndelta}}
\end{figure}

The momentum distributions of created EP pairs for two elliptically polarized electric fields with same-sign (left column) and opposite-sign (right column) ellipticity are calculated and shown in Fig. \ref{fig:pndelta}.
The ellipticity is chosen as $0.2$ (first row), $0.5$ (middle row), and $0.8$ (last row), respectively.
Other electric field parameters are $E_1=E_2=1.0\times10^7\mathrm{V}/\mathrm{m}$, $\omega=666.5\mathrm{THz}$, $\tau=9.0\mathrm{fs}$, and $T=6\tau$.
For $\delta_1=0.2$, the momentum distributions shown in \ref{fig:pndelta}(a) and (d) are extremely similar, which indicates that the electric fields in the $y$-direction have very minor influence on the momentum distribution. Furthermore, the signatures of  momentum distribution are also similar to the case where there is only $x$-direction electric fields.
For the case of $\delta_1=0.5$, the interfering fringes on momentum distribution shown in Fig. \ref{fig:pndelta}(b) are linked to form rings, while in Fig. \ref{fig:pndelta}(e), they are linked to form spirals.
This is significantly different from the case of EP pair production, where it was noted in Ref. \cite{Li2018hzi} that the spirals are sensitive to the ellipticity and will be broken over a wide range of ellipticity.
This may be related to the fact that there is a weaker exponential suppression in the EH pair production.
When $E_y$ is approximately equal to $0.5E_x$, the electric field in the $y$-direction with respect to the $x$-direction has been able to produce enough observable EH pairs.
This is expressed in the momentum distribution by filling in the region where the value of the distribution function is zero near $p_y=0$.
For the case of $\delta_1=0.8$, further homogenization of the momentum distribution can be seen in Figs. \ref{fig:pndelta}(c) and (f).
Although we only give results for partial ellipticity, it can be predicted that as the ellipticity increases from $0$ to $1$, the momentum distribution will gradually transform from the linearly polarized case to the circularly polarized case.

\section{CONCLUSIONS AND DISCUSSIONS}
\label{sec:five}
In this paper, the signatures of momentum distributions of EH pairs produced in graphene for two arbitrarily polarized electric field with or without a time delay are investigated using the low energy approximation model and compared with the case of EP pair production.

Firstly, we have studied the momentum distribution of created EH pairs for a single elliptically polarized electric field in the region of multi-photon resonance absorption.
The momentum distribution is very similar to that in EP pair production.
When the polarization value of the electric field $\delta=0$, the node structures on the momentum distribution can be well explained by the expression similar to the one given in EP pair production.
With the increase of the polarization value, the node structures gradually disappear, and homogeneous multi-photon absorption rings are formed on the momentum distribution.
Secondly, the momentum distribution of created EH pairs for two linearly polarized electric fields with a time delay and no field frequency is considered.
Unexpectedly, the momentum distribution for two co-directional electric field pulses has ring patterns in the $y$-direction, which is very different from that in EP pair production.
Thirdly, for the LLCP electric field, the momentum distribution exhibits the Ramsey interference fringes due to the phase difference between the amplitudes of created EH pairs for the two fields.
With the concept of turning points, a qualitative equation is given for determining the interference fringes.
Fourthly, for the LRCP electric field, the momentum distribution of created EH pairs has spiral structures that are more insensitive to the number of oscillation cycles of the electric field than that in EP pair production.
For a few-cycle electric field, the spiral structures of the momentum distribution is destroyed in EP pair production but not in the EH pair production.
Moreover, we have given an equation for qualitatively determining the positions of spirals, whose estimation is in good agreement with the numerical result.
Finally, we have found that for two elliptically polarized electric fields with a time delay, the interference pattern and spiral structures are much less affected by the ellipticity compared to the case in EP pair production.
For example, for the ellipticity $\delta_1=-\delta_2=0.5$, the momentum distribution has obvious spirals.

These results not only deepen our understanding of the EH pair production in graphene for complex electric fields but also show the similarities and differences between EH and EP pair production.
This demonstrates the justification for using graphene to simulate the EP pair production from vacuum.
Moreover, the electric field strength and frequency required for EH pair production are much easier to realize in the laboratory, which is beneficial to our understanding of EP pair production through solid-state systems in the future.

\begin{acknowledgments}
The work is supported by the National Natural Science Foundation of China (NSFC) under Grants No. 11974419 and No. 11705278, and by the Fundamental Research Funds for the Central Universities (No. 2023ZKPYL02).
\end{acknowledgments}

\end{document}